\documentclass[dvips]{arxstspdf}
\usepackage{graphicx}
\usepackage{flushend}
\usepackage{stfloats}

\volume{25}
\issue{4}
\pubyear{2010}
\firstpage{492}
\lastpage{505}
\doi{10.1214/08-STS264}

\begin{document}
\begin{frontmatter}

\title{The MM Alternative to EM}
\runtitle{The MM Alternative to EM}

\begin{aug}
\author[a]{\fnms{Tong Tong} \snm{Wu}} \and
\author[b]{\fnms{Kenneth} \snm{Lange}\corref{}\ead[label=e2]{klange@ucla.edu}}
\runauthor{T. T. Wu and K. Lange}

\affiliation{University of Maryland and University of California}

\address[a]{Tong Tong Wu is Assistant Professor,
Department of Epidemiology and Biostatistics, University of Maryland,
College Park, Maryland 20742, USA.}
\address[b]{Kenneth Lange is Professor,
Departments of Biomathematics, Human Genetics and Statistics,
University of California, Los Angeles, California 90095-1766, USA
\printead{e2}.}

\end{aug}

%
\begin{abstract}
The EM algorithm is a special case of a more general algorithm called the
MM algorithm. Specific MM algorithms often have nothing to do with
missing data. The first M step of an MM algorithm creates a surrogate
function that is optimized in the second M step. In minimization,
MM stands for majorize--minimize; in maximization, it stands for
minorize--maximize.
This two-step process always drives the objective function in the right
direction.
Construction of MM algorithms relies on recognizing and manipulating
inequalities rather than calculating conditional expectations. This
survey walks the reader through the construction of several specific
MM algorithms. The potential of the MM algorithm in solving high-dimensional
optimization and estimation problems is its most attractive feature.
Our applications to random graph models, discriminant analysis
and image restoration showcase this ability.
\end{abstract}

%
\begin{keyword}
\kwd{Iterative majorization}
\kwd{maximum likelihood}
\kwd{inequalities}
\kwd{penalization}.
\end{keyword}

\end{frontmatter}

\section{Introduction}\label{section1}

This survey paper tells a tale of two algorithms born in the same year.
We celebrate the christening of the EM algorithm by Dempster, Laird and
Rubin
(\citeyear{dempster77})
for good reasons. The EM algorithm is one of the workhorses of computational
statistics with literally thousands of applications. Its value was almost
immediately recognized by the international statistics community. The~more
general MM algorithm languished in obscurity for years. Although in
1970 the
numerical analysts Ortega and Rheinboldt (\citeyear{ortega70}) allude to the MM principle
in the context of line search methods, the first statistical
application occurs
in two papers (de Leeuw, \citeyear{deleeuw76}; de Leeuw and Heiser, \citeyear{deleeuw77})
of de Leeuw and Heiser in 1977 on multidimensional
scaling. One can argue that the unfortunate neglect of the de Leeuw and Heiser
papers has retarded the growth of computational statistics. The purpose of
the present paper is to draw attention to the MM algorithm and highlight
some of its interesting applications.

Neither the EM nor the MM algorithm is a concre\-te algorithm. They are
both principles for creating algorithms. The MM principle is based on
the notion
of (tangent) majorization. A function $g(\theta\mid\theta^{n})$ is
said to majorize a function $f(\theta)$ provided
%
\begin{eqnarray}\label{majorization_definition}
f(\theta^{n}) & = & g(\theta^{n} \vert\theta^{n}), \nonumber \\[-6pt]\\[-6pt]
f(\theta) & \le& g(\theta\vert\theta^{n}), \quad\theta
\ne\theta^{n}. \nonumber
\end{eqnarray}
In other words, the surface $\theta\mapsto g(\theta\vert\theta^{n})$
lies above the
surface $f(\theta)$ and is tangent to it at the point $\theta=\theta^{n}$.
Here $\theta^{n}$ represents the current iterate in a search of the surface
$f(\theta)$. The function $g(\theta\vert\theta^{n})$ minorizes
$f(\theta)$
if $-g(\theta\vert\theta^{n})$ majorizes $-f(\theta)$. Readers should
take heed that the term majorization is used in a different sense in
the theory of convex functions (Marshall and Olkin, \citeyear{marshall79}).

In the minimization version of the MM algorithm, we minimize the surrogate
majorizing function\break $g(\theta\vert\theta^{n})$ rather than the actual
function $f(\theta)$.
If $\theta^{n+1}$ denotes the minimum of the surrogate $g(\theta\vert
\theta^{n})$, then
one can show that the MM procedure forces $f(\theta)$ downhill.
Indeed, the relations
%
\begin{equation} \label{descent_property}
\quad f(\theta^{n+1})\le g(\theta^{n+1} \vert\theta^{n})
\le g(\theta^{n} \vert\theta^{n})_n= f(\theta^{n})
\end{equation}
follow directly from the definition of $\theta^{n+1}$ and the
majorization conditions
(\ref{majorization_definition}). The descent property (\ref
{descent_property}) lends
the MM algorithm remarkable numerical stability. Strictly speaking, it
depends only
on decreasing the surrogate function $g(\theta\vert\theta^{n})$, not
on minimizing it.
This fact has practical consequences when the minimum of $g(\theta\vert
\theta^{n})$
cannot be found exactly. In the maximization version of the MM algorithm,
we maximize the surrogate minorizing function $g(\theta\vert\theta^{n})$.
Thus, the acronym MM does double duty, serving as an abbreviation of both
pairs ``majorize--minimize'' and ``minorize--maximize.'' The earlier, less
memorable name
``iterative majorization'' for the MM algorithm unfortunately suggests that
the principle is limited to minimization.

The EM algorithm is actually a special case of the MM algorithm. If
$f(\theta)$
is the log-likelihood of the observed data, and $Q(\theta\vert\theta
^n)$ is
the function created in the E step, then the minorization
\[
f(\theta)  \ge Q(\theta\vert\theta^n)+f(\theta^n)-Q(\theta^n
\vert\theta^n)
\]
is the key to the EM algorithm. Maximizing \mbox{$Q(\theta\mid\theta^n)$}
with respect
to $\theta$ drives $f(\theta)$ uphill. The proof of the EM
minorization relies on
the nonnegativity of the Kullback--Leibler divergence of two conditional
probability
densities. The divergence inequality in turn depends on Jensen's
inequality and the
concavity of the function $\ln x$ (Hunter and Lange, \citeyear{hunter04b}; Lange,
\citeyear{lange04}).

In our opinion, the MM principle is easier to state and grasp than the
EM principle.
It requires neither a likelihood model nor a missing data framework. In
some cases,
existing EM algorithms can be derived more easily by isolating a key
majorization
or minorization. In other cases, it is quicker and more transparent to postulate
the complete data and calculate the conditional expectations required
by the E step
of the EM algorithm. Many problems involving the multivariate normal
distribution
fall into this latter category. Finally, EM and MM algorithms constructed
for the same problem can differ. Our second example illustrates this point.
Which algorithm is preferred is then a matter of reliability in finding
the global optimum, ease of implementation, speed of convergence
and computational complexity.

This is not the first survey paper on the MM algorithm and probably
will not
be the last. The previous articles (Becker, Yang and Lange, \citeyear{becker97};
de Leeuw, \citeyear{deleeuw94}; Heiser, \citeyear{heiser95}; Hunter and
Lange, \citeyear{hunter04b}; Lange, Hunter and Yang, \citeyear{lange00})
state the general principle, sketch various methods of majorization
and present a variety of new and old applications. Prior to these
survey papers,
the MM principle surfaced in robust regression (Huber, \citeyear{huber81}),
correspondence analysis (Heiser, \citeyear{heiser87}), the quadratic lower bound principle
(Bohning and Lindsay, \citeyear{bohning88}), alternating least squares applications
(Bijleveld and de Leeuw, \citeyear{bijleveld91}; Kiers, \citeyear{kiers02}; Kiers and Ten
Berge, \citeyear{kiers92}; Takane, Young and de Leeuw, \citeyear{takane77}),
medical imaging (De Pierro, \citeyear{depierro95}; Lange and Fessler, \citeyear{lange95})
and convex programming (Lange, \citeyear{lange94}). Recent work has demonstrated the
utility of MM algorithms in a broad range of statistical contexts, including
quantile regression (Hunter and Lange, \citeyear{hunter00}), survival analysis (Hunter and Lange,
\citeyear{hunter02}),
nonnegative matrix factorization (Eld\'{e}n, \citeyear{elden07}; Lee and Seung, \citeyear{lee99},
\citeyear{lee01};
Pauca, Piper and Plemmous, \citeyear{pauca06}),
paired and multiple comparisons (Hunter,\break\citeyear{hunter04a}), variable selection
(Hunter and Li, \citeyear{hunter05}), DNA sequence analysis (Sabatti and Lange,
\citeyear{sabatti02}) and
discriminant analysis (Groenen, Nalbantov and Bioch, \citeyear{groenen06}; Lange and Wu,
\citeyear{lange07}).

The primary purpose of this paper is to present MM algorithms not
featured in
previous surveys. Some of these algorithms are novel, and some are minor
variations on previous themes. Except for our first two examples in Sections
\ref{section2} and \ref{section3}, it is unclear whether any of the
algorithms can
be derived from a missing data perspective. This fact alone
distinguishes them
from standard EM fare. In digesting the examples, readers should notice
how the MM algorithm
interdigitates with other algorithms such as block relaxation and
Newton's method.
Classroom expositions of computational statistics leave the impression
that different optimization algorithms act in isolation. In reality,
some of
the best algorithms are hybrids. The examples also stress penalized estimation
and high-dimensional problems that challenge traditional algorithms such
as scoring and Newton's method. Such problems are apt to dominate computational
statistics and data mining for some time to come. The MM principle offers
a foothold in the unforgiving terrain of large data sets and high-dimensional
models.

Two theoretical skills are necessary for constructing new MM algorithms.
One is a good knowledge of statistical models. Another is proficiency with
inequalities. Most inequalities are manifestations of convexity. The
single richest
source of minorizations is the supporting hyperplane inequality
\begin{eqnarray*}
f(x) & \ge& f(y)+df(y)(x-y)
\end{eqnarray*}
satisfied by a convex function $f(x)$ at each point $y$ of its domain.
Here $df(y)$ is the row vector of partial derivatives of $f(x)$ at
$y$.

The quadratic lower bound principle of Bohning and Lindsay
(\citeyear{bohning88}) propels majorization when the objective function has
bounded curvature. Let $d^2f(x)$ be the second differential (Hessian) of
the objective function $f(x)$, and suppose $B$ is a positive definite matrix
such that $B-d^2f(x)$ is positive semidefinite for all arguments $x$. Then
we have the majorization
\begin{eqnarray*}
f(x) & = & f(y)+df(y)(x-y)\\
&&{}+\tfrac{1}{2}(x-y)^t d^2f(z)(x-y) \\
& \le& f(y)+df(y)(x-y)\\
&&{}+\tfrac{1}{2}(x-y)^t B(x-y),
\end{eqnarray*}
where $z$ falls on the line segment between $x$ and $y$. Minimization
of the
quadratic surrogate is straightforward. In the unconstrained
case, it involves inversion of the matrix $B$, but this can be done once
in contrast to the repeated matrix inversions of Newton's method.
Other relevant majorizations and minorizations will be mentioned
as needed. Readers wondering where to start in brushing up on
inequalities are urged to consult the elementary exposition
(Steele, \citeyear{steele04}). The more advanced texts (Boyd and Vandenberghe,
\citeyear{boyd04}; Lange,
\citeyear{lange04})
are also useful for statisticians.

Finally, let us stress that neither EM nor MM is a panacea.
Optimization is
as much art as science. There is no universal algorithm of choice, and a
good deal of experimentation is often required to choose among EM, MM, scoring,
Newton's method, quasi-Newton methods, conjugate gradient, and other
more exotic
algorithms. The simplicity of MM algorithms usually argues in their
favor. Balanced
against this advantage is the sad fact that many MM algorithms exhibit
excruciatingly
slow rates of convergence. Section \ref{section8} derives the theoretical criterion governing
the rate of convergence of an MM algorithm. Fortunately, MM algorithms are
readily amenable to acceleration. For the sake of brevity, we will omit a
detailed development of acceleration and other important topics. Our discussion
in Section \ref{section9} will take these up and point out pertinent
references.

\section{Estimation with the Multivariate~$t$}\label{section2}

The multivariate $t$-distribution has density
\begin{eqnarray*}
f(x) & = & {\Gamma\biggl({\nu+p \over2}\biggr)}\\
&&{}\cdot\biggl\{\Gamma\biggl({\nu\over2}\biggr)(\nu\pi)^{p/2}
|\Omega|^{1/2}\\
&&\quad {}\cdot\biggl[1+\frac{1}{\nu}({ x}-{\mu})^t{\Omega}^{-1}({
x}-{\mu})\biggr]^{(\nu+p)/2}\biggr\}^{-1}
\end{eqnarray*}
for all $x \in{\mathsf R}^p$. Here $\mu$ is the mean vector, $\Omega
$ is the positive
definite scale matrix and $\nu>0$ is the degrees of freedom. Let
$x_1,\ldots,x_m$ be a
random sample from $f(x)$. To estimate $\mu$ and $\Omega$ for $\nu$
fixed, the well-known
EM algorithm (Lange, Little and Taylor, \citeyear{lange89}; Little and Rubin,
\citeyear{little02}) iterates according to
%
\begin{eqnarray}
\mu^{n+1} & = & {1 \over s^n} \sum_{i=1}^m w^{n}_{i} x_i, \label{mean_update} \\
\Omega^{n+1} & = & {1 \over m} \sum_{i=1}^m w^{n}_{i} (x_i-\mu
^{n+1})(x_i-\mu^{n+1})^t,
\label{variance_update}
\end{eqnarray}
where $s^{n} = \sum_{i=1}^m w^{n}_{i}$ is the sum of the case weights
\begin{eqnarray*}
w^{n}_{i} & = & {\nu+p \over\nu+ d^{n}_{i}}, \quad
d^{n}_{i} = (x_i-\mu^n)^t (\Omega^n)^{-1}(x_i-\mu^n).
\end{eqnarray*}
The derivation of the EM algorithm hinges on the representation of the
$t$-density
as a hidden mixture of multivariate normal densities.

Derivation of the same algorithm from the MM perspective ignores
the missing data and exploits the concavity of the function $\ln x$. Thus,
the supporting hyperplane inequality
\begin{eqnarray*}
-\ln x & \ge& - \ln y - {x-y \over y}
\end{eqnarray*}
implies the minorization
\begin{eqnarray*}
& & -{1 \over2}\ln|\Omega| -{\nu+p \over2} \ln[\nu+(x_i-\mu)^t
\Omega^{-1}(x_i-\mu)] \\
&&\quad  \ge -{1 \over2}\ln|\Omega|\\
&&\qquad {} - {\nu+p \over2} \biggl[ \ln{\nu+p
\over w^{n}_{i}}\\
&&\hspace*{67pt}{}+ \bigl(\nu+(x_i-\mu)^t \Omega^{-1}(x_i-\mu)\\
&& \hspace*{139pt}{}-(\nu+p)/w^{n}_{i}\bigr)\\
&&\hspace*{133pt}{}\cdot\bigl((\nu+p)/w^{n}_{i} \bigr)^{-1} \biggr] \\
&&\quad  =  -{1 \over2}\ln|\Omega| - {w^{n}_{i} \over2}[\nu+(x_i-\mu)^t
\Omega^{-1}(x_i-\mu)]\\
&&\qquad {}+c_i^{n}
\end{eqnarray*}
for case $i$, where $c_i^{n}$ is a constant that depends on neither
$\mu$ nor $\Omega$. Summing
over the different cases produces the overall surrogate. Derivation of
the updates
(\ref{mean_update}) and (\ref{variance_update}) reduces to standard
manipulations with
the multivariate normal (Lange, \citeyear{lange04}).

Kent, Tyler and Vardi (\citeyear{kent94}) suggest an alternative algorithm that replaces
the EM update (\ref{variance_update}) for $\Omega$ by
%
\begin{eqnarray}
\Omega^{n+1} & = & {1 \over s^{n}} \sum_{i=1}^m w^{n}_{i}
(x_i-\mu^{n+1})(x_i-\mu^{n+1})^t . \label{revised_variance_update}
\end{eqnarray}
Megan and van Dyk (\citeyear{meng97}) justify this modest amendment by expanding the
parameter space to include a working parameter that is tweaked to produce faster
convergence. It is interesting that a trivial variation of our
minorization produces the Kent, Tyler and Vardi (\citeyear{kent94}). We simply combine
the two log terms and minorize via
\begin{eqnarray*}
& & -{1 \over2}\ln|\Omega| -{\nu+p \over2} \ln[\nu+(x_i-\mu)^t
\Omega^{-1}(x_i-\mu)] \\
& &\quad{} =  -{\nu+p \over2} \ln\{|\Omega|^a[\nu+(x_i-\mu)^t \Omega
^{-1}(x_i-\mu)]\} \\
& &\quad{} \ge -{w^{n}_{i} \over2|\Omega^n|^a}\{|\Omega|^a[\nu+(x_i-\mu
)^t \Omega^{-1}(x_i-\mu)]\}\\
&&\qquad {}+c_i^{n} ,
\end{eqnarray*}
with working parameter $a=1/(\nu+p)$.

For readers wanting the full story, we now indicate briefly how the
second step of the MM
algorithm is derived. This revolves around maximizing the surrogate function
\begin{eqnarray*}
-\sum_{i=1}^m w^{n}_{i}\{|\Omega|^a[\nu+(x_i-\mu)^t \Omega
^{-1}(x_i-\mu)]\}
\end{eqnarray*}
with respect to $\mu$ and $\Omega$. Regardless of the value of~
$\Omega$, one should
choose $\mu$ as the weighted mean (\ref{mean_update}). If we let $R$
be the square root of
$\Omega^{n+1}$ as defined by (\ref{revised_variance_update})
and substitute $\mu^{n+1}$
in the surrogate, then the refined surrogate function can be expressed
\begin{eqnarray*}
&&-s^{n} \{|\Omega|^a[\nu+ \operatorname{tr}(\Omega^{-1}R^2)]\}\\
&&\quad  =  -s^{n} \{ |R^{-1}\Omega R^{-1}|^a [\nu+\operatorname
{tr}(R\Omega^{-1}R)]\}|R|^{2a}.
\end{eqnarray*}
To show that $\Omega= R^2$ minimizes the surrogate, let $\lambda
_1,\ldots,\lambda_p$
denote the eigenvalues of the positive definite matrix $R^{-1} \Omega R^{-1}$.
This allows us to express the surrogate as a negative multiple of the function
\begin{eqnarray*}
h(\lambda) & = & \nu\prod_{j=1}^p \lambda_j^a + \prod_{j=1}^p
\lambda_j^a \sum_{j=1}^p \lambda_j^{-1}.
\end{eqnarray*}
The choice $\lambda={\bf1}$ corresponds to $\Omega= R^2$ and yields
the value
$h({\bf1}) = \nu+p$. The identity $\Omega= R^2$ can now be proved by showing
that $\nu+p$ is a lower bound for $h(\lambda)$. Setting $\lambda_j =
e^{\theta_j}$, a simple
rearrangement of the bounding inequality shows that it suffices to
prove the alternative
inequality
\begin{eqnarray*}
e^{-{1 /(\nu+p)}\sum_{j=1}^p \theta_j} & \le& {\nu\over\nu
+p}e^0 +{1 \over\nu+p} \sum_{j=1}^p e^{-\theta_j},
\end{eqnarray*}
which is a direct consequence of the convexity of $e^x$.

\section{Grouped Exponential Data} \label{section3}

The EM algorithm for estimating the intensity of grouped exponential
data is well known (Dempster, Laird and Rubin, \citeyear{dempster77}; McLachlan and
Krishnan, \citeyear{mclachlan97}; Meilijson, \citeyear{meilijson89}). In this
setting the complete data corresponds to a random sample $x_1,\ldots
,x_m$ from
an exponential density with intensity $\lambda$. The observed data conforms
to a sequence of thresholds $t_1<t_2 < \cdots< t_m$. It is convenient
to append the threshold $t_0=0$ to this list and to let $c_i$ record
the number of values that fall within the interval $(t_i,t_{i+1}]$. The
exceptional count $c_m$ represents the number of right-censored values.
One can derive a novel MM algorithm by close examination of the log-likelihood
\begin{eqnarray*}
L(\lambda) & = & c_0 \ln(1-e^{-\lambda t_1})\\
&&{}+\sum_{i=1}^{m-1} c_i \ln(e^{-\lambda t_i}-e^{-\lambda t_{i+1}}) -
c_m \lambda t_m \\
& = & - \lambda\sum_{i=0}^{m-1} c_i t_{i+1} - c_m \lambda t_m
+ \sum_{i=0}^{m-1} c_i \ln(e^{\lambda d_i}-1) ,
\end{eqnarray*}
where $d_i = t_{i+1}-t_i$.

\begin{table*}[t]
\tabcolsep=0pt
\caption{Comparison of MM and EM on grouped exponential data}\label{table3}
\begin{tabular*}{\textwidth}{@{\extracolsep{\fill}}lcccc@{}}
 \hline
& \multicolumn{2}{c}{ \textbf{MM algorithm}} & \multicolumn{2}{c@{}}{ \textbf{EM algorithm}}\\
\ccline{2-3,4-5}
{$\bolds{n}$} & {$\bolds{\lambda^n}$} & {$\bolds{L(\lambda^n)}$} & {$\bolds{\lambda^n}$} &
{$\bolds{L(\lambda^n)}$}\\ \hline
0 & 1.00000 & $-$3.00991 & 1.00000 & $-$3.00991 \\
1 & 0.50000 & $-$1.75014 & 0.27082 & $-$1.34637 \\
2 & 0.25000 & $-$1.32698 & 0.21113 & $-$1.30591 \\
3 & 0.18924 & $-$1.30528 & 0.20102 & $-$1.30443 \\
4 & 0.19762 & $-$1.30438 & 0.19904 & $-$1.30437 \\
5 & 0.19848 & $-$1.30437 & 0.19864 & $-$1.30437 \\
6 & 0.19853 & $-$1.30437 & 0.19856 & $-$1.30437 \\
7 & 0.19854 & $-$1.30437 & 0.19854 & $-$1.30437 \\ \hline
\end{tabular*}
\end{table*}

The above partial linearization of the log-likelihood $L(\lambda)$ focuses
our attention on the remaining nonlinear parts of $L(\lambda)$
determined by the function $f(\lambda) = \ln(e^{\lambda d}-1)$.
The derivatives
\begin{eqnarray*}
f'(\lambda) & = & {e^{\lambda d}d \over e^{\lambda d}-1}, \quad
f''(\lambda) =  - {e^{\lambda d}d^2 \over(e^{\lambda d}-1)^2}
\end{eqnarray*}
indicate that $f(\lambda)$ is increasing and concave. It is impossible
to minorize $f(\lambda)$ by a linear function, so we turn to the
quadratic lower bound principle. Hence, in the second-order Taylor expansion
\begin{eqnarray*}
f(\lambda) & = & f(\lambda^n)+f'(\lambda^n)(\lambda-\lambda^n)\\
&&{}+\tfrac{1}{2} f''(\mu)(\lambda-\lambda^n)^2 ,
\end{eqnarray*}
with $\mu$ between $\lambda$ and $\lambda^n$, we seek to
bound $f''(\mu)$ from below. One can easily check that $f''(\mu)$
is increa\-sing on $(0,\infty)$ and tends to $-\infty$ as $\mu$
approaches~$0$. To avoid this troublesome limit, we restrict~$\lambda$
to the interval $({1 \over2}\lambda^n,\infty)$ and substitute
$f''({1 \over2}\lambda^n)$ for $f''(\mu)$. Minorizing the nonlinear
part of $L(\lambda)$ term by term now gives a quadratic minorizer
$q(\lambda)$
of $L(\lambda)$. Because the coefficient of $\lambda^2$ in $q(\lambda)$
is negative, the restricted maximum $\lambda^{n+1}$ of $q(\lambda)$ occurs
at the boundary ${1 \over2}\lambda^n$ whenever the unrestricted maximum
occurs to the left of ${1 \over2}\lambda^n$. In symbols, the MM update
reduces to
\begin{eqnarray*}
\lambda^{n+1} & = & \max\biggl\{{1 \over2}\lambda^n,\\
&&\hspace*{26pt}{}\lambda^n +{\sum_{i=0}^{m-1} c_i(v^n_i-t_{i+1})-c_mt_m \over\sum_{i=0}^{m-1}
c_i w^n_i}\biggr\} ,
\end{eqnarray*}
where
\begin{eqnarray*}
v^n_i & = & {e^{\lambda^n d_i}d_i \over e^{\lambda^n d_i}-1}, \quad
w^n_i =  {e^{\lambda^n d_i/2}d_i^2/4 \over(e^{\lambda^n
d_i/2}-1)^2} .
\end{eqnarray*}
Table \ref{table3} compares the MM algorithm and the traditional EM
algorithm on the toy example of Meilijson (\citeyear{meilijson89}). Here we have
$m = 3$ thresholds at 1, 3 and 10 and assign proportions $0.185$, $0.266$,
$0.410$ and $0.139$ to the four ordinal groups. It is clear that the
MM algorithm hits its lower bound on iterations 1 and 2. Although its local
rate of convergence appears slightly better than that of the EM
algorithm, the
differences are minor. The purpose of this exercise is more to illustrate
the quadratic lower bound principle in deriving MM algorithms.

\section{Power Series Distributions} \label{section4}

\begin{table*}[t]
\caption{Performance of the algorithm (\protect\ref{power_series_iteration})
for truncated Poisson data}\label{table1}
\begin{tabular*}{\textwidth}{@{\extracolsep{\fill}}lccccc@{}}
 \hline
{$\bolds{n}$} & {$\bolds{\theta^n}$} & {$\bolds{L(\theta^n)}$} & {$\bolds{n}$} &
{$\bolds{\theta^n}$} & {$\bolds{L(\theta^n)}$} \\
\hline
0 & 1.00000 & $-$5.41325 & \phantom{0}7 & 1.59161 & $-$4.34467 \\
1 & 1.26424 & $-$4.63379 & \phantom{0}8 & 1.59280 & $-$4.34466 \\
2 & 1.43509 & $-$4.40703 & \phantom{0}9 & 1.59329 & $-$4.34466 \\
3 & 1.52381 & $-$4.35635 & 10 & 1.59349 & $-$4.34466 \\
4 & 1.56424 & $-$4.34670 & 11 & 1.59357 & $-$4.34466 \\
5 & 1.58151 & $-$4.34501 & 12 & 1.59360 & $-$4.34466 \\
6 & 1.58867 & $-$4.34472 & 13 & 1.59362 & $-$4.34466 \\
\hline
\end{tabular*}
\end{table*}

A family of discrete density functions $p_k(\theta)$ defined on
$\{0,1,\ldots\}$ and indexed by a parameter $\theta> 0$ is said
to be a power series family provided for all $k$
%
\begin{eqnarray}
p_k(\theta) & = & \frac{c_k \theta^k}{q(\theta)} , \label{powerden}
\end{eqnarray}
where $c_k \geq0$ and $q(\theta) = \sum_{k=0}^\infty c_k \theta^k$
is the
appropriate normalizing constant (Rao, \citeyear{rao73}). The binomial, negative
binomial,
Poisson and logarithmic families are examples. Zero truncated versions
of these families also qualify. Fisher scoring is the traditional approach
to maximum likelihood estimation with a power series family. If
$x_1,\ldots,x_m$
is a random sample from the discrete density (\ref{powerden}), then
the log-likelihood
\begin{eqnarray*}
L(\theta) & = & \sum_{i=1}^m x_i \ln\theta- m \ln q(\theta)
\end{eqnarray*}
has score $s(\theta)$ and expected information $J(\theta)$
\begin{eqnarray*}
s(\theta) & = & {1 \over\theta} \sum_{i=1}^m x_i - {m q'(\theta)
\over q(\theta)} ,
\quad J(\theta)  = \frac{m \sigma^2(\theta)}{\theta^2},
\end{eqnarray*}
where $\sigma^2(\theta)$ is the variance of a single realization.

Functional iteration provides an alternative to scoring. It is clear
that the maximum likelihood estimate $\hat{\theta}$ is a root of the equation
%
\begin{eqnarray}
\bar{x} & = & \frac{\theta q'(\theta)}{q(\theta)} \label{moment_estimate},
\end{eqnarray}
where $\bar{x}$ is the sample mean. This result suggests the iteration scheme
%
\begin{eqnarray}
\theta^{n+1} & = & {\bar{x} q(\theta^n) \over q'(\theta^n)} =
 M(\theta^n)
\label{power_series_iteration}
\end{eqnarray}
and raises two obvious questions. First, is the algorithm (\ref
{power_series_iteration})
an MM algorithm? Second, is it likely to converge to $\hat{\theta}$
even in
the absence of such a guarantee? Local convergence hinges on the
derivative condition
$|M'(\hat{\theta})|<1$. When this condition holds, the map $\theta
^{n+1}=M(\theta^n)$
is locally contractive near the fixed point $\hat{\theta}$. It turns out
that
\begin{eqnarray*}
M'(\hat{\theta}) & = & 1 - {\sigma^2(\hat{\theta}) \over\mu(\hat
{\theta})},
\end{eqnarray*}
where
\begin{eqnarray*}
\mu(\theta) & = & {\theta q'(\theta) \over q(\theta)}
\end{eqnarray*}
is the mean of a single realization $X$. Thus, convergence depends on the
ratio of the variance to the mean. To prove these assertions it is helpful
to differentiate~$q(\theta)$. The first derivative delivers the mean
and the second derivative the second factorial moment
\begin{eqnarray*}
\mathrm{E}[X(X-1)] & = & {\theta^2 q''(\theta) \over q(\theta)}.
\end{eqnarray*}
If one substitutes these into the obvious expression for $M'(\hat
{\theta})$ and
invokes equality (\ref{moment_estimate}) at $\hat{\theta}$, then the
moment form
of $M'(\hat{\theta})$ emerges.

\begin{table*}[t]
\caption{Performance of the algorithm (\protect\ref{power_series_iteration})
for logarithmic data \label{table2}}
\begin{tabular*}{\textwidth}{@{\extracolsep{\fill}}lccccc@{}}
\hline
{$\bolds{n}$} & {$\bolds{\theta^n}$} & {$\bolds{L(\theta^n)}$} & {$\bolds{n}$} &
{$\bolds{\theta^n}$} & {$\bolds{L(\theta^n)}$} \\
\hline
0 & 0.99000 & $-$15.47280 & \phantom{0}9 & 0.71470 & $-$8.98294 \\
1 & 0.09210 & $-$24.32767 & 10 & 0.71565 & $-$8.98293 \\
2 & 0.17545 & $-$18.35307 & 11 & 0.71517 & $-$8.98293 \\
3 & 0.31814 & $-$13.30624 & 12 & 0.71542 & $-$8.98293 \\
4 & 0.52221 & $-$9.96349 & 13 & 0.71529 & $-$8.98293 \\
5 & 0.70578 & $-$8.98560 & 14 & 0.71535 & $-$8.98293 \\
6 & 0.71991 & $-$8.98355 & 15 & 0.71532 & $-$8.98293 \\
7 & 0.71291 & $-$8.98310 & 16 & 0.71534 & $-$8.98293 \\
8 & 0.71655 & $-$8.98297 & 17 & 0.71533 & $-$8.98293 \\
\hline
\end{tabular*}
\end{table*}

To address the question of whether functional iteration is an MM algorithm,
we make the assumption that $q(\theta)$ is log-concave. This condition holds
for the binomial and Poisson distributions but not for the negative binomial
and logarithmic distributions. The convexity of $-\ln q(\theta)$
entails the
minorization,
\begin{eqnarray*}
L(\theta) & \ge& \sum_{i=1}^m x_i \ln\theta- m \ln q(\theta^n) -
m [\ln q(\theta^n)]'(\theta-\theta^n) \\
& = & \sum_{i=1}^m x_i \ln\theta-m \ln q(\theta^n) - m {q'(\theta
^n) \over q(\theta^n)}(\theta-\theta^n).
\end{eqnarray*}
Setting the derivative of this surrogate function equal to~0 leads to the
MM update (\ref{power_series_iteration}). One can demonstrate that
log-concavity
implies $\sigma^2(\theta) \le\mu(\theta)$. The~local contraction condition
$|M'(\hat{\theta})|<1$ is consistent with the looser criterion
$\sigma^2(\theta) \le2 \mu(\theta)$. Thus, there is room for a
viable local
algorithm that fails to have the ascent property.

The truncated Poisson density has normalizing function $q(\theta) =
e^{\theta}-1$.
The second derivative test shows that $q(\theta)$ is log-concave.
Table \ref{table1}
records the well-behaved MM iterates (\ref{power_series_iteration})
for the
choices $\bar{x}=2$ and \mbox{$m=10$}. The geometric density counting
failures until a success
has normalizing function $q(\theta)=(1-\theta)^{-1}$, which is
log-convex rather
than log-concave. The iteration function is now $M(\theta) = \bar
{x}(1-\theta)$.
Since \mbox{$M'(\theta) = -\bar{x}$}, the algorithm diverges for $\bar{x}>1$.
Finally, the discrete logarithmic density has normalizing constant
$q(\theta) = - \ln(1-\theta)$,
which is also log-convex rather than log-concave. The choices $\bar
{x}=2$ and $m=10$
lead to the iterates in Table \ref{table2}. Although the algorithm~
(\ref{power_series_iteration})
converges for the logarithmic density, it cannot be an MM algorithm
because the log-likelihood
experiences a decline at its first iteration.

One of the morals of this example is that many natural algorithms
only satisfy the descent or ascent property in special circumstances.
This is
not necessarily a disaster, but without such a guarantee, safeguards must
usually be instituted to prevent iterates from going astray. Proof of
the descent or
ascent property almost always starts with majorization or minorization.
Because so much of statistical inference revolves around log-likelihoods,
log-convexity and log-concavity are possibly more important than
ordinary convexity and concavity in constructing MM algorithms.

There are a variety of criteria that help in checking log-concavity.
Besides the obvious second derivative test, one should
keep in mind the closure properties of the collection of log-concave
functions on a given domain (Bergstrom and Bagnoli, \citeyear{bergstrom05}; Boyd and
Vandenberghe, \citeyear{boyd04}). For example,
the collection is closed under the formation of products and positive powers.
Any positive concave function is log-concave. If $f(x) > \alpha\ge0$
for all
$x$, then $f(x)-\alpha$ is log-concave. In some cases, integration preserves
log-concavity. If $f(x)$ is log-concave, then $\int_a^x f(y)\,dy$ and
$\int_x^b f(y)\,dy$ are log-concave. When $f(x,y)$ is jointly log-concave
in $(x,y)$, $\int f(x,y)\, dy$ is log-concave in $x$. As a special case,
the convolution of two log-concave functions
is log-concave. One of the more useful recent tests for log-concavity
pertains to power series (Anderson, Vamanamurthy and Vuorinen,
\citeyear{anderson07}).\break
Suppose $f(x) = \sum_{k=0}^\infty a_k x^k$
has radius of convergence $r$ around the origin. If the coefficients
$a_k$ are
positive and the ratio $(k+1)a_{k+1}/a_k$ is decreasing in $k$, then $f(x)$
is log-concave on $(-r,r)$. This result also holds for finite series
$f(x) = \sum_{k=0}^m a_k x^k$. In minorization, log-convexity plays the
linearizing role of log-concavity. The closure properties of the set of
log-convex
functions are equally impressive (\cite{boyd04}).

\section{A Random Graph Model} \label{section5}

Random graphs provide interesting models of connectivity in genetics
and internet
node ranking. Here we consider the random graph model of
Blitzstein, Chatterjee and Diaconis (\citeyear{blitzstein08}). Their model assigns
a nonnegative propensity
$p_i$ to each node $i$. An edge between nodes $i$ and $j$ then forms
independently with
probability $p_ip_j/(1+p_ip_j)$. The most obvious statistical question
in the model
is how to estimate the $p_i$ from data. Once this is done, we can rank
nodes by
their estimated propensities.

If $E$ denotes the edge set of the graph, then the log-likelihood can be
written as
%
\begin{eqnarray}
L(p) & = & \sum_{\{i,j\} \in E} [\ln p_i + \ln
p_j]\nonumber\\[-6pt]\\[-6pt]
&&{}-\sum_{\{i,j\}} \ln(1+p_i p_j).\nonumber \label{graph_loglikelihood}
\end{eqnarray}
Here $\{i,j\}$ denotes a generic unordered pair. The logarithms $\ln(1+p_ip_j)$
are the bothersome terms in the log-likelihood. We will minorize each of
these by
exploiting the convexity of the function $-\ln(1+x)$. Application of the
supporting hyperplane inequality yields
\begin{eqnarray*}
-\ln(1+p_i p_j) & \ge& -\ln(1+p_{i}^n p_{j}^n)\\
&&{}-{1 \over1+p_{i}^n p_{j}^n}
(p_ip_j-p_{i}^n p_{j}^n)
\end{eqnarray*}
and eliminates the logarithm. Note that equality holds when $p_i =
p_{i}^n$ for
all $i$. This minorization is not quite good enough to separate
parameters, however.
Separation can be achieved by invoking the second minorizing inequality
\begin{eqnarray*}
-p_ip_j & \ge& -{1 \over2}\biggl({p_{j}^n \over p_{i}^n}p_i^2
+{p_{i}^n \over p_{j}^n}p_j^2\biggr) .
\end{eqnarray*}
Note again that equality holds when all $p_i=p_{i}^n$.

\begin{table*}[t]
\caption{Convergence of the MM random graph algorithm \label{table4}}
\begin{tabular*}{\textwidth}{@{\extracolsep{\fill}}lcccc@{}}
\hline
{$\bolds{n}$} & {$\bolds{p_{0}^n}$} & {$\bolds{p_{m/2}^n}$} & {$\bolds{p_{m}^n}$} &
{$\bolds{L(p^n)}$}\\
\hline
\phantom{0}0 & 0.00100 & 0.48240 & 0.95613 & $-$40572252.7109 \\
\phantom{0}1 & 0.00000 & 0.48281 & 0.97251 & $-$40565250.8333 \\
\phantom{0}2 & 0.00000 & 0.48220 & 0.98274 & $-$40562587.5350 \\
\phantom{0}3 & 0.00000 & 0.48151 & 0.98950 & $-$40561497.1411 \\
\phantom{0}4 & 0.00000 & 0.48093 & 0.99408 & $-$40561038.9534 \\
\phantom{0}5 & 0.00000 & 0.48050 & 0.99720 & $-$40560843.3998 \\
10 & 0.00000 & 0.47963 & 1.00299 & $-$40560695.6515 \\
15 & 0.00000 & 0.47950 & 1.00387 & $-$40560693.1245 \\
20 & 0.00000 & 0.47948 & 1.00400 & $-$40560693.0770 \\
25 & 0.00000 & 0.47948 & 1.00403 & $-$40560693.0761 \\
30 & 0.00000 & 0.47948 & 1.00403 & $-$40560693.0761 \\
35 & 0.00000 & 0.47948 & 1.00403 & $-$40560693.0764 \\
\hline
\end{tabular*}
\end{table*}

These considerations imply that up to a constant $L(p)$ is minorized by
the function
\begin{eqnarray*}
g(p \vert p^n) & = & \sum_{\{i,j\} \in E} [\ln p_i + \ln p_j]\\
&&{}- \sum_{\{i,j\}}{1 \over1+p_{i}^n p_{j}^n}{1 \over2}\biggl({p_{j}^n \over p_{i}^n}p_i^2
+{p_{i}^n \over p_{j}^n}p_{j}^2 \biggr).
\end{eqnarray*}
The fact that $g(p \vert p^n)$ separates parameters allows us\vspace*{2pt} to compute
$p_{i}^{n+1}$
by setting the derivative of $g(p \vert p^n)$ with respect to $p_i$
equal to 0.
Thus, we must solve
\begin{eqnarray*}
0 & = & \sum_{\{i,j\} \in E} {1 \over p_i} -
\sum_{j \ne i} {1 \over1+p_{i}^n p_{j}^n}{p_{j}^n \over p_{i}^n}p_i.
\end{eqnarray*}
If $d_i = \sum_{\{i,j\} \in E} 1$ denotes the degree of node $i$, then
the positive square root
%
\begin{eqnarray}
p_{i}^{n+1} & = & \biggl[{p_{i}^n d_i \over\sum_{j \ne i}
{p_{j}^n/(1+p_{i}^n p_{j}^n)}}\biggr]^{1/2}
\label{mmgraph}
\end{eqnarray}
is the pertinent solution. Blitzstein, Chatterjee and Diaconis (\citeyear{blitzstein08})
derive a
different and possibly more effective algorithm by a contraction mapping
argument.

The MM update (\ref{mmgraph}) is not particularly intuitive, but it
does have
the virtue of algebraic simplicity. When\vadjust{\goodbreak} $d_i = 0$, it also makes
the sensible choice $p_{i}^{n+1}=0$. As a check on our derivation,
observe that a stationary point of the log-likelihood satisfies
\begin{eqnarray*}
0 & = & {d_i \over p_i} - \sum_{j \ne i} {p_j \over1+p_i p_j} ,
\end{eqnarray*}
which is just a rearranged version of the update (\ref{mmgraph}) with iteration
superscripts suppressed.

The MM algorithm just derived carries with it certain guarantees. It is
certain to
increase the log-likelihood at every iteration, and if its maximum value
is attained
at a unique point, then it will also converge to that point. It is
straightforward
to prove that the log-likelihood is concave under the reparameterization
$p_i=e^{-q_i}$.
The~requirement of two successive minorizations in our derivation gives
us pause
because if minorization is not tight, then convergence is slow. On the
other hand,
if the number of nodes is large, then competing algorithms such as Newton's
method entail large matrix inversions and are very expensive.

As a test case for the MM algorithm, we generated a random graph on $m =10{,}000$
nodes with a propensity~$p_i$ for node $i$ of $(i-{1 \over2})/m$. To derive
appropriate starting values for the propensities, we estimated a common
background propensity $q$ by setting $q^2/(1+q^2)$ equal to the ratio of
observed edges to possible edges and solving for $q$. This background
propensity was
then used to estimate each $p_i$ by setting $p_iq/(1+p_iq)$ equal to
$d_i/m$ and
solving for $p_i$. Table \ref{table4} displays the components
$p_{0}^n$, $p_{m/2}^n$
and $p_{m}^n$ of the parameter vector $p^n$ at iteration $n$. The log-likelihood
actually fails the ascent test in the last iteration because its
rightmost digits
are beyond machine precision. Despite this minor flaw, the algorithm performs
impressively on this relatively large and decidedly nonsparse problem.
As an
indication of the quality of the final estimate $\hat{p}$, the maximum
error $\max_i |(i-{1 \over2})/m-\hat{p}_i|$ was 0.0825 and the
average absolute
error ${1 \over m}\sum_i |(i-{1 \over2})/m-\hat{p}_i|$ was 0.0104.

\section{Discriminant Analysis} \label{section6}

Discriminant analysis is another attractive application. In
discriminant analysis with two categories, each case $i$ is
characterized by a feature vector $z_i$ and a category membership
indicator $y_i$ taking the values $-1$ or $1$. In the machine
learning approach to discriminant analysis (Scholkopf and Smola, \citeyear{scholkopf02};
Vapnik, \citeyear{vapnik95}),
the hinge loss function $[1-y_i(\alpha+z_i^t \beta)]_+$ plays a
prominent role. Here $(u)_+$ is shorthand for the convex function
$\max\{u,0\}$. Just as in ordinary regression, we can penalize the
overall loss
\begin{eqnarray*}
g(\theta) & = & \sum_{i=1}^n [1-y_i(\alpha+z_i^t \beta)]_+
\end{eqnarray*}
by imposing a lasso or ridge penalty (Hastie, Tibshirani and Friedman, \citeyear{hastie01}).
Note that the linear
regression function $h_i(\theta) = \alpha+z_i^t \beta$ predicts
either $-1$ or $1$.
If $y_i = 1$ and $h_i(\theta)$ overpredicts in the sense that
$h_i(\theta)>1$,
then there is no loss. Similarly, if $y_i=-1$ and $h_i(\theta)$ underpredicts
in the sense that $h_i(\theta)<-1$, then there is no loss.

\begin{table*}[t]
\caption{Empirical examples from UCI machine learning repository\label{table5}}
\begin{tabular*}{\textwidth}{@{\extracolsep{\fill}}lcccccc@{}}
 \hline
\textbf{Data set} & \multicolumn{3}{c}{\textbf{Hinge loss}} & \multicolumn{3}{c@{}}
{\textbf{VDA}}\\[-6pt]
\multicolumn{1}{@{}c}{\hrulefill} & \multicolumn{3}{c}{\hrulefill} & \multicolumn{3}{c@{}}
{\hrulefill}\\
(\textbf{Cases, features}) & \textbf{Iters} & \textbf{Error} & \textbf{Time} & \textbf{Iters} &
\textbf{Error} & \textbf{Time} \\
\hline
Diabetes (768, 8) & \phantom{0}44 & 0.2266 & 0.063 & 11 & 0.2240 & 0.015 \\
SPECT (80, 22) & 326 & 0.2000 & 0.359 & \phantom{0}7 & 0.1750 & 0.000 \\
Tic-tac-toe (958, 9) & 274 & 0.0167 & 0.578 & 26 & 0.0167 & 0.062 \\
Ionosphere (351, 33) & 483 & 0.0513 & 2.984 & 42 & 0.0570 & 0.266 \\
\hline
\end{tabular*}
\end{table*}

Most strategies for estimating $\theta$ pass to the dual of
the original minimization problem. A simpler strategy
is to majorize each contribution to the loss by a quadratic and minimize
the surrogate loss plus penalty. A little calculus
(Groenen, Nalbantov and Bioch, \citeyear{groenen06}) shows that $(u)_+$ is majorized
at $u^n \ne0$
by the quadratic
%
\begin{eqnarray}
q(u \vert u^n) & = & {1 \over4|u^n|}(u+|u^n|)^2 . \label
{hinge_majorizer}
\end{eqnarray}
In fact, this is the best quadratic majorizer (\cite{deleeuw07}).
To avoid the singularity at 0, we recommend replacing $q(u \mid u^n)$ by
\begin{eqnarray*}
r(u \vert u^n) & = & {1 \over4|u^n| + \varepsilon}(u+|u^n|)^2.
\end{eqnarray*}
In double precision, a good choice of $\varepsilon$ is $10^{-5}$.
If we impose a ridge penalty, then the majorization (\ref{hinge_majorizer})
leads to a pure MM algorithm exploiting weighted least squares.

If the number of predictors is large, then the matrix inversions
entailed in updating all parameters simultaneously become burdensome.
Coordinate descent offers a viable alternative because it updates
a single parameter at a time. The large number of iterations until
convergence required by coordinate descent is often outweighed by the extreme
simplicity of each parameter update. Quadratic majorization of the hinge
losses keeps the updates simple and guarantees a reduction in the
objective function. The decisions to use a lasso or ridge penalty and
apply pure MM or coordinate descent with majorization will be dictated in
practical problems by considerations of model selection and the
number of potential predictors.

In discriminant analysis with more than two categories,
it is convenient to pass to $\varepsilon$-insensitive loss
and multiple linear regression. Our recently introduced method
of vertex discriminant analysis (VDA) (Lange and Wu, \citeyear{lange07}) operates in this
fashion and relies on an MM algorithm. If there are $k+1$ categories
and $p$ predictors, the basic idea is situate the class indicators at
the vertices of a regular simplex in $\mathsf{R}^k$ and minimize the criterion
%
\begin{eqnarray}
R(A,b) & = & {1 \over n} \sum_{i=1}^n \|y_i-A z_i
-b\|_{\varepsilon}\\
&&{}+ \lambda\sum_{j=1}^k \|a_j\|^2,\nonumber \label{discriminant_objective}
\end{eqnarray}
where $y_i$ is the vertex assigned to case $i$, $a_j^t$ is the
$j$th row of a $k \times p$ matrix $A$ of regression coefficients, $b$
is a $k \times1$ column vector of
intercepts, and
%
\begin{eqnarray}
\|v\|_{\varepsilon} & = & \max\{\|v\|-\varepsilon,0\} \label{e-distance}
\end{eqnarray}
is $\varepsilon$-insensitive Euclidean distance. Once $A$ and $b$ are estimated,
we can assign a new case to the closest vertex, and hence category.
One can design a quadratic surrogate by application of the Cauchy--Schwarz
inequality and minimize the surrogate by solving $k$ coordinated least squares
problems. The combination of a parsimonious loss function and an efficient
MM algorithm make VDA one of the most effective discriminant analysis methods
tested (Lange and Wu, \citeyear{lange07}).

As a comparison of hinge-loss discriminant analysis versus VDA, we
now consider four typical examples from the UCI machine learning
repository (Asuncion and Newman, \citeyear{uci}). All four examples involve just two categories.
For each data set, Table \ref{table5} lists the numbers of
cases, features, and iterations until convergence, as well as the
training error rates
and the computing times in seconds under both hinge loss and \mbox{$\varepsilon
$-insensitive}
loss. For VDA we set $\varepsilon=0.9999$, just below the recommended cutoff
of $\sqrt{(2k+2)/k}/2=1$ for $k+1=2$ categories. The cutoff is the
largest $\varepsilon$ avoiding overlap of the \mbox{$\varepsilon$-insensitive} spheres
around each vertex of the regular simplex. We chose the value $10^{-2}$
for the tuning parameter $\lambda$ in all four examples. Our previous numerical
experience shows that VDA is relatively insensitive to the choice of
$\lambda$.
Inspection of the training errors suggests that the two methods have
similar accuracy.
To our surprise, VDA is considerably faster.

\section{Image Restoration and Inpainting} \label{section7}

The MM algorithm is also employed in image deconvolution (Bioucas-Dias, Figueiredo and Oliveira,
\citeyear{bioucas-dias06}; Liao et al., \citeyear{liao02}).
Suppose a photograph is divided into pixels and $y_{ij}$ is the
digitized intensity
for pixel $(i,j)$. Some of the $y_{ij}$ are missing or corrupted. Smoothing
pixel values can give a visually improved image. Correction of pixels
subject to minor corruption is termed denoising; correction of missing
or grossly
distorted values is termed inpainting. Let $S$ be the set of pixels with
acceptable values. We can restore the photograph by minimizing the criterion
\begin{eqnarray*}
\sum_{(i,j) \in S} (y_{ij} - \mu_{ij})^2 + \lambda\sum_i \sum_j
\sum_{(k,l) \in N_{ij}} \|\mu_{ij}-\mu_{kl}\|_{\mathrm{TV}} ,
\end{eqnarray*}
where $N_{ij}$ denotes the pixels neighboring pixel $(i,j)$,
$\|x\|_{\mathrm{TV}} = \sqrt{x^2+\varepsilon}$ is the total variation norm
with \mbox{$\varepsilon>0$} small and $\lambda>0$ is a tuning constant.
Let $\mu_{ij}^n$ be the current iterate. The total variation
penalties are majorized using
\begin{eqnarray*}
\|x\|_{\mathrm{TV}} & \le& \|x^n\|_{\mathrm{TV}}+{1 \over2 \|x^n\|_{\mathrm{TV}}}[x^2-(x^n)^2]
\end{eqnarray*}
based on the concavity of the function $\sqrt{t+\varepsilon}$. These maneuvers
construct a simple surrogate function expressible as a weighted sum of squares.
Other roughness penalties are possible. For instance, the scaled sum of squares
$ \lambda\sum_i \sum_j \sum_{(k,l) \in N_{ij}} (\mu_{ij}-\mu_{kl})^2$
is plausible. Unfortunately, this choice tends to deter the formation
of image edges. The total variation alternative is preferred in practice
because it is gentler while remaining continuously differentiable.

If the pixels are defined on a rectangular grid, then we can divide
them into two blocks in a checkerboard fashion, with the red checkerboard
squares falling into one block and the black checkerboard squares into
the other block. Within a block, the least squares problems generated by
the surrogate function are parameter separated and hence trivial to solve.
Thus, it makes sense to alternate the updates of the blocks. Within a block
we update $\mu_{ij}$ via
\begin{eqnarray*}
\mu_{ij}^{n+1} & = & \frac{2y_{ij}+ \lambda\sum_{(k,l) \in N_{ij}}
{\mu_{kl}^n}/{\| \mu_{ij}^{n}-\mu_{kl}^{n}\|_{\mathrm{TV}}}}
{2 + \lambda\sum_{(k,l) \in N_{ij}} {1}/{\| \mu_{ij}^{n}-\mu
_{kl}^{n}\|_{\mathrm{TV}}}}
\end{eqnarray*}
for $(i,j) \in S$ or via
\begin{eqnarray*}
\mu_{ij}^{n+1} & = & \frac{\sum_{(k,l) \in N_{ij}}
{\mu_{kl}^n}/{\| \mu_{ij}^{n}-\mu_{kl}^{n}\|_{\mathrm{TV}}}}
{\sum_{(k,l) \in N_{ij}}{1}/{\| \mu_{ij}^{n}-\mu_{kl}^{n}\|
_{\mathrm{TV}}}}
\end{eqnarray*}
for $(i,j) \notin S$. Here each interior pixel $(i,j)$ has four neighbors.
If the singularity constant $\varepsilon$ is too small or if the tuning
$\lambda$
is too large, then small residuals generate very large weights. When this
pitfall is avoided, the described algorithm is apt to be superior to the
fused lasso algorithm of Friedman, Hastie and Tibshirani (\citeyear{friedman07}).

\begin{figure*}

\includegraphics{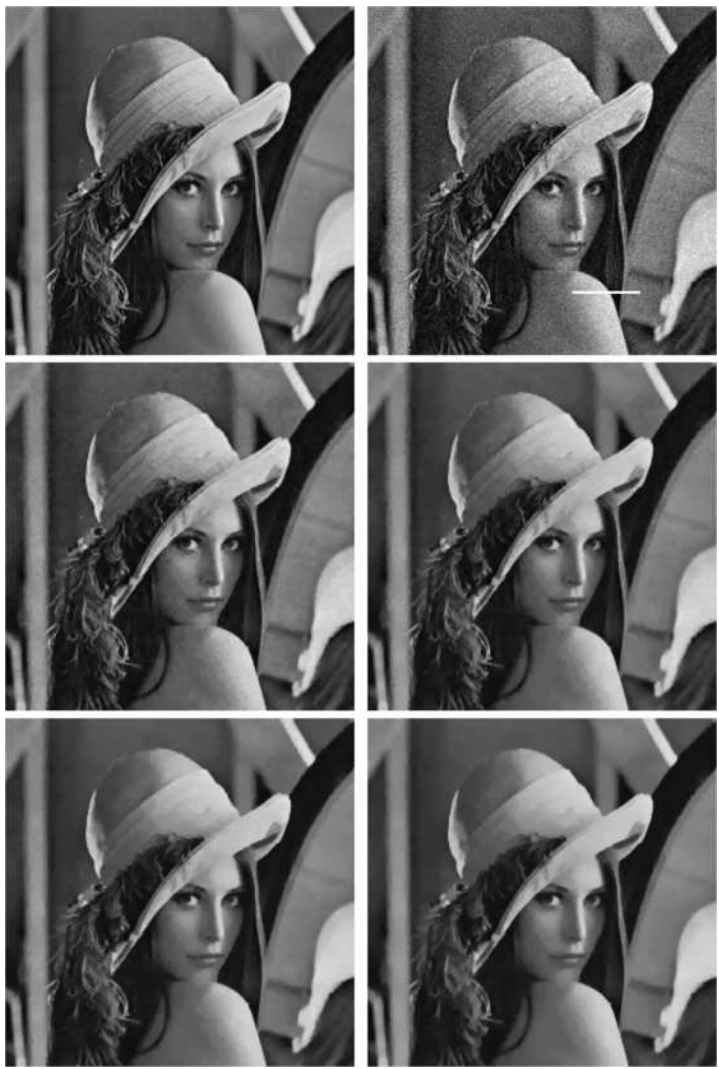}

\caption{Restoration of the Lenna photograph.  Top row left: the original image;
top row right: image corrupted by Gaussian noise (mean 0 and standard deviation 10) and
a scratch; second row left: restored image with $\lambda=10$; second row right: restored
image with $\lambda=15$; third row left: restored image with $\lambda=20$; third row
right: restored image with $\lambda=25$.  The same value $\varepsilon = 1$ is used
throughout.}\label{inpainting_lenna_stddev10}
\end{figure*}

We applied the total variation algorithm to the standard image of the model
Lenna. Figure \ref{inpainting_lenna_stddev10} shows the original
$256 \times256$ image with pixel values digitized on a gray scale from
0 to
255. To the right of the original image is a version corrupted by
Gaussian noise
(mean 0 and standard deviation 10) and a scratch on the shoulder. The
images are restored
with $\lambda$ values of 10, 15, 20 and 25 and an $\varepsilon$ value
of 1. Although
we tend to prefer the restoration on the right in the second row, this
is a
matter of judgment. Variations in $\lambda$ clearly control the
balance between
image smoothness and loss of detail.

\section{Local Convergence of MM Algorithms} \label{section8}

Many MM and EM algorithms exhibit a slow rate of convergence. How can one
predict the speed of convergence of an MM algorithm and choose between competing
algorithms? Consider an MM map $M(\theta)$ for minimizing the
objective function
$f(\theta)$ via the surrogate function $g(\theta\vert\theta^n)$.
According to a theorem of Ortega (\citeyear{ortega90}), the local rate of convergence
of the sequence $\theta^{n+1}=M(\theta^n)$ is determined by the
spectral radius
$\rho$ of the differential $dM(\theta^\infty)$ at the minimum point
$\theta^\infty$
of $f(\theta)$. Well-known calculations (Dempster, Laird and Rubin, \citeyear{dempster77};
\cite{lange95a}) demonstrate that
\begin{eqnarray*}
dM(\theta^\infty) & = & I - d^{2}g(\theta^\infty\vert\theta^\infty
)^{-1} d^2f(\theta^\infty).
\end{eqnarray*}
Hence, the eigenvalue equation $dM(\theta^\infty)v = \lambda v$ can be
rewritten as
\begin{eqnarray*}
d^{2}g(\theta^\infty\vert\theta^\infty)v - d^2f(\theta^\infty)v
& = & \lambda d^{2}g(\theta^\infty\vert\theta^\infty)v.
\end{eqnarray*}
Taking the inner product of this with $v$, we can solve for $\lambda$
in the form
\begin{eqnarray*}
\lambda& = & 1 - {v^t d^2f(\theta^\infty)v \over v^t d^{2}g(\theta
^\infty\vert\theta^\infty)v}.
\end{eqnarray*}
Extension of this line of reasoning shows that the spectral radius
satisfies
\begin{eqnarray*}
\rho& = & 1 - \min_{v \ne{\bf0}} {v^t d^2f(\theta^\infty)v
\over v^t d^{2}g(\theta^\infty\vert\theta^\infty)v} .
\end{eqnarray*}
Thus, the rate of convergence of the MM iterates is determined by how
well $d^{2}g(\theta^\infty\vert\theta^\infty)$ approximates
$d^{2}f(\theta^\infty)$. In practice, the surrogate function
$g(\theta\vert\theta^n)$
should hug $f(\theta)$ is tightly as possible for $\theta$ close to~$\theta^n$.

Meng and van Dyk (\citeyear{meng97}) use this Rayleigh quotient characterization
of the spectral radius to prove that the Kent et al.\ multivariate $t$
algorithm is faster
than the original multivariate $t$ algorithm. In essence, they show
that the second
differential $d^{2}g(\theta\vert\theta)$ is uniformly more positive
definite for the
alternative algorithm. de Leeuw and Lange (\citeyear{deleeuw07}) make substantial
progress in designing optimal quadratic surrogates. For most other MM
algorithms, however,
such theoretical calculations are too hard to carry out, and one must
rely on numerical
experimentation to determine the rate of convergence. The uncertainties
about rates of
convergence are reminiscent of the uncertainties surrounding MCMC
methods. This should
not deter us from constructing MM algorithms. On large-scale problems,
many traditional
algorithms are simply infeasible. If we can construct a MM algorithm,
then there is
always the chance of accelerating it. We take up this topic briefly in
the discussion.
Finally, let us stress that the number of iterations until convergence is
not the sole determinant of algorithm speed. Computational complexity per
iteration also comes into play. On this basis, a standard MM algorithm
for transmission
tomography is superior to a plausible but different EM algorithm (Lange,
\citeyear{lange04}).

\section{Discussion} \label{section9}

Perhaps the best evidence of the pervasive influence of the EM algorithm
is the sheer number of citations garnered by the Dempster et al. paper. As of April 2008,
Google Scholar lists 11,232 citations.
By contrast, Google Scholar lists 58 citations for the de Leeuw
 paper and 47 citations for the de Leeuw and Heiser paper.
If our contention about the relative importance of the EM and MM
algorithms is true, how can one account for this disparity? Several
reasons come to mind. One is the venue of publication. The~\textit{Journal
of the
Royal Statistical Society, Series B}, is one of the most widely read journals
in statistics. The de Leeuw and Heiser papers are buried in a hard to access
conference proceedings. Another reason is the prestige of the authors. Four
of the five authors of the three papers, Nan Laird, Donald Rubin, Jan
de Leeuw
and Willem Heiser, were quite junior in 1977. On the other hand, Arthur Dempster
was a major figure in statistics and well established at Harvard, the most
famous American university. Besides these extrinsic differences, the
papers have
intrinsic differences that account for the better reception of the
Dempster et al. paper.
Its most striking advantage is the breadth of its subject matter.
Dempster et al.
were able to unify different branches of computational statistics under
the banner of
a clearly enunciated general principle. de Leeuw and Heiser stuck to
multidimensional
scaling. Their work and extensions are well summarized by Borg and Groenen (\citeyear{borg97}).

The EM algorithm immediately appealed to the stochastic intuition of
statisticians,
who are good at calculating the conditional expectations required by
the E step.
The MM algorithm relies on inequalities and does not play as well to the
strengths of statisticians. Partly for this reason the MM algorithm had
difficulty breaking out of the vast but placid backwater of social science
applications where it started. It remained sequestered there for years, nurtured
by several highly productive Dutch statisticians with less clout than
their American
and British colleagues.

Our emphasis on concrete applications neglects some issues of considerable
theoretical and practical importance. The most prominent of these are
global convergence analysis, computation of asymptotic standard errors,
acceleration, and approximate solution of the optimization step (second
M) of the
MM algorithm. Let us address each of these in turn.

Virtually all of the convergence results announced by
Dempster et al. (\citeyear{dempster77}) and corrected by Wu (\citeyear{wu83}) and Boyles
(\citeyear{boyles83})
carry over to the MM algorithm. The known theory, both local and
global, is
summarized in the references (Lange, \citeyear{lange04}; Vaida, \citeyear{vaida05}). As
anticipated, the best results hold in the presence of convexity or concavity.
The SEM algorithm of Meng and Rubin (\citeyear{meng91}) for computation of
asymptotic standard errors also carries over to the MM algorithm
(Hunter, \citeyear{hunter04a}). Numerical differentiation of the score function
is a viable competitor, particularly if the score can be
evaluated analytically. The simplest form of acceleration is step
doubling (de Leeuw and Heiser, \citeyear{deleeuw80}; Lange and Fessler, \citeyear{lange95}).
This maneuver replaces the point
delivered by an algorithm map $\theta^{n+1} = M(\theta^{n})$ by
the new point $\theta^{n}+2[M(\theta^{n})-\theta^{n}]$. Step doubling
usually halves the number of iterations until convergence in an
MM algorithm. More effective forms of acceleration are possible using
matrix polynomial extrapolation (Varadhan and Roland, \citeyear{varadhan08}) and
quasi-Newton and conjugate gradient elaborations of the MM algorithm
(Jamshidian and Jennrich, \citeyear{jamshidian97}; Lange, \citeyear{lange95b}). Finally,
if the optimization step
of an MM algorithm cannot be accomplished analytically, it is possible
to fall back on the MM gradient algorithm (Hunter and Lange, \citeyear{hunter04b}; Lange,
\citeyear{lange95a}).
Here one substitutes one step of Newton's method for full optimization
of the surrogate function $g(\theta\vert\theta^{n})$ with respect
to $\theta$. Fortunately, this approximate algorithm has exactly the
same rate of convergence as the original MM algorithm. It also
preserves the descent or ascent property of the MM algorithm close to the
optimal point.\looseness=-1

The reader may be left wondering whether EM or MM
provides a clearer path to the derivation of new algorithms. In the
absence of a likelihood function, it is difficult for EM to work its
magic. Even so, criteria such as least squares can involve hidden
likelihoods. Perhaps the best reply is that we are asking the
wrong question. After all, one man's mathematical meat is often
another man's mathematical poison. A better question is whether MM
broadens the possibilities for devising new algorithms. In our
view, the answer to the second question is a resounding yes. Our
last four examples illustrate this point. Of course, it may be
possible to derive one or more of these algorithms from the EM perspective,
but we have not been clever enough to do so.

In highlighting the more general MM algorithm, we intend no disrespect
to the pioneers of the EM algorithm. If the fog of obscurity lifts from
the MM algorithm, it will not detract from their achievements. It
may, however, propel the ambitious plans for data mining underway in
the 21st century. Even with the expected advances in computer
hardware, the statistics community still needs to concentrate on
effective algorithms. The MM principle is poised to claim a share
of the credit in this enterprise. Statisticians with a numerical
bent are well advised to add it to their toolkits.

\section*{Acknowledgment}
Research supported in part by USPHS Grants\break
GM53275 and MH59490 to KL.

\vspace*{-2pt}
\end{document}